# Achievable Rates for the Multiple Access Channel with Feedback and Correlated Sources

Lawrence Ong and Mehul Motani

Department of Electrical & Computer Engineering
National University of Singapore
{lawrence.ong,motani}@nus.edu.sg

**Abstract**

In this paper, we investigate achievable rates on the multiple access channel with feedback and correlated sources (MACFCS). The motivation for studying the MACFCS stems from the fact that in a sensor network, sensors collect and transmit correlated data to a common sink. We derive two achievable rate regions for the three-node MACFCS.

## 1 Introduction

We consider a sensor network in which every sensor is capable of transmitting as well as receiving. Each sensor collects data and aims to send them to a single destination. We note that the data collected by the sensor nodes might be correlated, e.g., if they are located close to one another. Taking into account these facts, we model the two-sensor single-sink network by the channel depicted in Fig. 1. We term this channel the *multiple access channel with feedback and correlated sources* (MACFCS).

This channel is a combination of the multiple access channel with correlated sources (MACCS) and the multiple access channel with feedback (MACF). The MACCS (with a common part) was studied by Slepian and Wolf [1], who derived an achievable rate region. In their paper, separate source coding and channel coding are used, where source coding is first performed to remove the correlation between the two sources and then channel coding for the multiple access channel (MAC) with independent sources is employed. The MACCS (with possibly no common part) was considered by Cover *et al.* [2]. They showed, by using a simple example, that separating source and channel coding is not optimal and derived an achievable rate region for the MACCS.

The MACF (with independent sources) was investigated by Cover and Leung [3]. In their model, there are two sources and all nodes, i.e., the two sources and the destination, receive the same channel output. A year later, Carleial [4] further generalized the channel to the case where each node receives a different channel output signal.

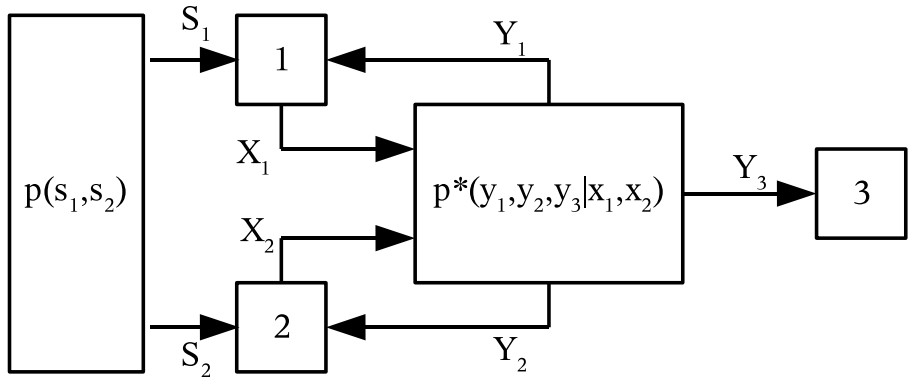

Figure 1: The three-node multiple access channel with feedback and correlated sources.

The three-node discrete memoryless MACFCS is denoted by $\big(\mathcal{S}_1 \times \mathcal{S}_2, p(s_1, s_2), \mathcal{X}_1 \times \mathcal{X}_2, p^*(y_1, y_2, y_3|x_1, x_2), \mathcal{Y}_1 \times \mathcal{Y}_2 \times \mathcal{Y}_3\big)$. $s_1 \in \mathcal{S}_1$ and $s_2 \in \mathcal{S}_2$ are the source messages to node 1 and 2 respectively and they are drawn from the discrete bivariate distribution $p(s_1, s_2)$. Here, $\mathcal{S}_1, \mathcal{S}_2, \mathcal{X}_1, \mathcal{X}_2, \mathcal{Y}_1, \mathcal{Y}_2$, and $\mathcal{Y}_3$ are seven finite sets. $p^*(y_1, y_2, y_3|x_1, x_2)$ defines the channel transition probability on $\mathcal{Y}_1 \times \mathcal{Y}_2 \times \mathcal{Y}_3$ for each $(x_1, x_2) \in \mathcal{X}_1 \times \mathcal{X}_2$. $x_1$ and $x_2$ are the inputs to the channel from nodes 1 and 2 respectively. $y_1$, $y_2$, and $y_3$ are the channel outputs to nodes 1, 2, and the destination respectively. We say that $(s_1, s_2)$ can be reliably transmitted to the destination if the probability that the destination wrongly decodes a pair of $(\mathbf{s}_1, \mathbf{s}_2) \in \mathcal{S}_1^n \times \mathcal{S}_2^n$ in each $n$ channel uses can be made arbitrarily small, for large $n$.

To the best of our knowledge, only Murugan *et al.* [5] have considered such a channel. However, they only considered Gaussian channels. Their approach is based on joint source-channel coding using time division multiple access (TDMA). Our work differs from [5] in that arbitrary channels (not only Gaussian channels) are considered. In addition, we consider the case where the source nodes can transmit and receive at the same time, meaning we do not restrict the transmission scheme to TDMA.

We classify coding strategies for the MACFCS into two categories. We now describe these strategies in more detail.

## 1.1 Full Decoding at Sources

For full decoding at sources, the general idea is for the sources to communicate so that each source has full information about what the other sources have. They then cooperate to send the combined signal to the destination. A scheme was proposed in [5], where the transmissions are split into two phases. In the first phase, the source nodes communicate with each other using TDMA. At the end of the first phase, each source has full information of what all other sources have. In the second phase, all sources cooperate to transmit to the destination. In this paper we offer an alternative solution (see Section 2) for full decoding at sources. Each source node transmits cooperative information of the previous block (which it decodes from other nodes) and new information (which is to be decoded by other sources and the destination) simultaneously. Since all nodes agree on the same fully decoded information of the previous block, *coherent combining* can be achieved. Under certain channel conditions, that all nodes fully decode the information of all other nodes might not be desirable. One example is when node 1 is far from the destination and node 2 is near to the destination. In this case, it is not necessary for node 1 to decode node 2's information. This leads us to the second strategy, in which full decoding is only done at the destination.

## 1.2 Full Decoding at Destination

For full decoding at the destination, source coding is first performed at every source node. This does not require physical communication among the sources. From [1], each source node performs source coding and forms independent inputs to the channel encoder. This removes the correlation between the sources.

At this point, we have turned the problem into channel coding for the MACF with independent sources. An achievable region of the MACF was obtained by Carleial [4]. We call that the partial decode-forward strategy. In this paper, we find another achievable region for the MACF using the compress-forward strategy (see Section 3.3). Combining the rate constraints of the source coding (for correlated sources) and the channel coding (for the MACF), we arrive at other achievable rate regions for the MACFCS. To the best of our knowledge, the compress-forward strategy has not been studied on the MACF.

## 2 Full Decoding at Sources

In this strategy, every node decodes the information from all other nodes and they cooperate to send information to the destination. We note that for the nodes to cooperate, they must first agree on the messages. In order to do this, they must first decode transmissions from the other nodes. We consider the following correlation structure in which the sources have a common part to send. [1] Let $D, E$, and $F$ be three independent random variables, equi-probable in $\{1, 2, \ldots, d\} = \mathcal{D}, \{1, 2, \ldots, e\} = \mathcal{E}$, and $\{1, 2, \ldots, f\} = \mathcal{F}$ respectively. Node 1 receives $S_1 = (D, E) \in \mathcal{D} \times \mathcal{E}$ and node 2 receives $S_2 = (D, F) \in \mathcal{D} \times \mathcal{F}$.

We note that node 1 does not know $F$ and node 2 does not know $E$. The idea here is for node 1 to decode $F$ (from node 2's transmission) and for node 2 to decode $E$. After decoding, both nodes have the full information $(D, E, F)$. They cooperate to send the fully decoded information as well as new information that is unknown to and to be decoded by other nodes. In summary, node 1 sends $(E, D', E', F')$ and node 2 sends $(F, D', E', F')$, where prime denotes the previous block's information.

---
[1] We assumed here that sources have a common part to send for the purpose of illustration only. The analysis in this section applies equally well to arbitrarily correlated sources.

## 2.1 An Achievable Rate Region

**Theorem 1** *Let $(\mathcal{S}_1 \times \mathcal{S}_2, p(s_1, s_2), \mathcal{X}_1 \times \mathcal{X}_2, p^*(y_1, y_2, y_3|x_1, x_2), \mathcal{Y}_1 \times \mathcal{Y}_2 \times \mathcal{Y}_3)$ be a discrete memoryless three-node MACFCS. $(s_1, s_2)$ can be reliably transmitted to the destination if the following holds.*

$$H(S_1|S_2) < \min[I(X_1; Y_2|W_0, W_1, W_2, X_2), I(W_1; Y_3|W_0, W_2) + I(X_1; Y_3|W_0, W_1, W_2, X_2)] \tag{1a}$$

$$H(S_2|S_1) < \min[I(X_2; Y_1|W_0, W_1, W_2, X_1), I(W_2; Y_3|W_0, W_1) + I(X_2; Y_3|W_0, W_1, W_2, X_1)] \tag{1b}$$

$$I(S_1; S_2) < I(W_0; Y_3|W_1, W_2), \tag{1c}$$

$$H(S_1) < I(W_0, W_1; Y_3|W_2) + I(X_1; Y_3|W_0, W_1, W_2, X_2), \tag{1d}$$

$$H(S_2) < I(W_0, W_2; Y_3|W_1) + I(X_2; Y_3|W_0, W_1, W_2, X_1), \tag{1e}$$

$$H(S_1|S_2) + H(S_2|S_1) < I(W_1, W_2; Y_3|W_0) + I(X_1, X_2; Y_3|W_0, W_1, W_2), \tag{1f}$$

$$H(S_1, S_2) < I(X_1, X_2; Y_3), \tag{1g}$$

*where* $p(x_1, x_2, y_1, y_2, y_3, w_0, w_1, w_2) = p(w_0)p(w_1)p(w_2)p(x_1|w_0, w_1, w_2)p(x_2|w_0, w_1, w_2) \times p^*(y_1, y_2, y_3|x_1, x_2, x_3)$. $W_0$, $W_1$ and $W_2$ are auxiliary random variables.

In the next section, we give a brief outline of the proof for Theorem 1.

## 2.2 Encoding and Decoding

First, we describe the coding scheme. Using Slepian and Wolf's Theorem 2 in [6], we know that when node 1 only knows $S_1 = (D, E)$ and node 2 knows $S_2 = (D, F)$, node 1 can encode $E$ using $H(S_1|S_2)$ bits and it can decoded by node 2. Similarly, node 2 can use $H(S_2|S_1)$ bits to encode $F$. The codebook generation is as follows:

1. Fix the probability mass functions $p(w_0), p(w_1), p(w_2), p(x_1|w_0, w_1, w_2)$, and $p(x_2|w_0, w_1, w_2)$.

2. Generate $2^{n[I(S_1;S_2)+\epsilon]}$ i.i.d. sequences $\mathbf{w_0}$ according to $\prod_{i=1}^n p(w_{0i})$. Index them $\mathbf{w}_0(i)$, $i \in \{1, 2, \ldots, 2^{n[I(S_1;S_2)+\epsilon]}\}$.

3. Generate $2^{n[H(S_1|S_2)+\epsilon]}$ i.i.d. sequences $\mathbf{w_1}$ according to $\prod_{i=1}^n p(w_{1i})$. Index them $\mathbf{w}_1(j)$, $j \in \{1, 2, \ldots, 2^{n[H(S_1|S_2)+\epsilon]}\}$.

4. Generate $2^{n[H(S_2|S_1)+\epsilon]}$ i.i.d. sequences $\mathbf{w_2}$ according to $\prod_{i=1}^n p(w_{2i})$. Index them $\mathbf{w}_2(k)$, $k \in \{1, 2, \ldots, 2^{n[H(S_2|S_1)+\epsilon]}\}$.

5. Define $h' = (i', j', k')$. For each $(\mathbf{w}_0(i'), \mathbf{w}_1(j'), \mathbf{w}_2(k'))$, generate $2^{n[H(S_1|S_2)+\epsilon]}$ sequences $\mathbf{x}_1$ according to $\prod_{i=1}^n p(x_{1i}|w_{0i}(i'), w_{1i}(j'), w_{2i}(k'))$. Index them $\mathbf{x}_1(j, h')$, $j \in \{1, 2, \ldots, 2^{n[H(S_1|S_2)+\epsilon]}\}$.

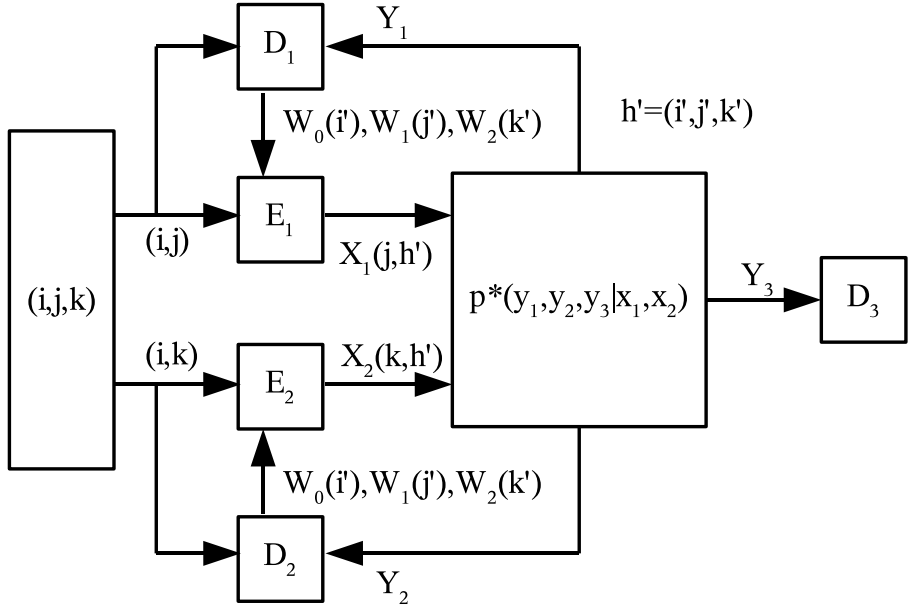

Figure 2: Coding for the multiple access channel with feedback and correlated sources using the decode-forward strategy.

6. Again for each $(\mathbf{w}_0(i'), \mathbf{w}_1(j'), \mathbf{w}_2(k'))$, independently generate $2^{n[H(S_2|S_1)+\epsilon]}$ sequences $\mathbf{x}_2$ according to $\prod_{i=1}^{n} p(x_{2i}|w_{0i}(i'), w_{1i}(j'), w_{2i}(k'))$.
Index them $\mathbf{x}_2(k, h')$, $k \in \{1, 2, \ldots, 2^{n[H(S_2|S_1)+\epsilon]}\}$.

The encoding steps (refer to Fig. 2) are as follows:

1. Assume that node 1 correctly estimates the index $k'$ sent by node 2 in the previous block. Using its information of $(i', j')$ from the previous block, it determines $h' = (i', j', k')$. Here, we use prime to indicate the index from the previous block. Observing a new block of $n$ input symbols $\mathbf{s}_1 \in (\mathcal{D} \times \mathcal{E})^n$, node 1 selects $j$ to represent $\mathcal{E}^n$. It can find a unique $j$ with probability tending to 1 if $j$ is encoded with no less than $n(H(S_1|S_2) + \epsilon)$ bits [1]. It then transmits $\mathbf{x}_1(j, h'_1)$.

2. Similarly, assuming that node 2 correctly decodes $j'$, it determines $h' = (i', j', k')$. It transmits $\mathbf{x}_2(k, h'_2)$. It can find a unique $k$ with probability tending to 1 if $k$ is encoded with no less than $n(H(S_2|S_1) + \epsilon)$ bits [1].

The decoding steps are as follows:

1. Upon observing the sequence $\mathbf{y}_1$, node 1 declares $\hat{k}$ has been sent by node 2 if there exists a unique $\hat{k}$ such that $\left(\mathbf{x}_1(j, h'), \mathbf{w}_0(i'), \mathbf{w}_1(j'), \mathbf{w}_2(k'), \mathbf{x}_2(\hat{k}, h'), \mathbf{y}_1\right) \in \mathcal{A}_\epsilon$. We use hat to indicate the estimate. Here, $\mathcal{A}_\epsilon$ is the set of jointly typical sequences (pg. 195 in [7]). We note that node 1 knows $h' = (i', j', k')$, which is the full information from the previous block, and its own information $j$. It can determine the correct $k$ with diminishing error probability if

$$H(S_2|S_1) < I(X_2; Y_1|W_0, W_1, W_2, X_1). \tag{2}$$

2. Similarly, observing the sequence $\mathbf{y}_2$, node 2 declares $\hat{j}$ has been sent by node 1 if there exists a unique $\hat{j}$ such that $\left(\mathbf{x}_1(\hat{j}, h'), \mathbf{w}_0(i'), \mathbf{w}_1(j'), \mathbf{w}_2(k')\mathbf{x}_2(k, h'), \mathbf{y}_2\right) \in \mathcal{A}_\epsilon$. Node 2 can determine the correct $j$ with diminishing error probability if

$$H(S_1|S_2) < I(X_1; Y_2|W_0, W_1, W_2, X_2). \tag{3}$$

3. The destination (node 3) decodes $(\hat{i}, \hat{j}, \hat{k})$ over two blocks. In the first block, assuming that it has already correctly decoded $h' = (i', j', k')$ from the previous block, it finds a set of $(\hat{j}, \hat{k}) \in \mathcal{L}_1$ where $\left(\mathbf{x}_1(\hat{j}, h'), \mathbf{x}_2(\hat{k}, h'), \mathbf{w}_0(i'), \mathbf{w}_1(j'), \mathbf{w}_2(k'), \mathbf{y}_3\right) \in \mathcal{A}_\epsilon$. In the second block, it then finds another set of $(\hat{j}, \hat{k}) \in \mathcal{L}_2$ and a unique $\hat{i}$ where $\left(\mathbf{w}_0(\hat{i}), \mathbf{w}_1(\hat{j}), \mathbf{w}_2(\hat{k}), \mathbf{y}_3\right) \in \mathcal{A}_\epsilon$. It declares $(\hat{i}, \hat{j}, \hat{k})$ has been sent if there is a unique $\hat{i}$ and a unique pair of $(\hat{j}, \hat{k})$ in $\mathcal{L}_1 \cap \mathcal{L}_2$. This can be done with diminishing error probability if

$$I(S_1; S_2) < I(W_0; Y_3|W_1, W_2), \tag{4a}$$
$$H(S_1|S_2) < I(W_1; Y_3|W_0, W_2) + I(X_1; Y_3|W_0, W_1, W_2, X_2), \tag{4b}$$
$$H(S_2|S_1) < I(W_2; Y_3|W_0, W_1) + I(X_2; Y_3|W_0, W_1, W_2, X_1), \tag{4c}$$
$$H(S_1) < I(W_0, W_1; Y_3|W_2) + I(X_1; Y_3|W_0, W_1, W_2, X_2), \tag{4d}$$
$$H(S_2) < I(W_0, W_2; Y_3|W_1) + I(X_2; Y_3|W_0, W_1, W_2, X_1), \tag{4e}$$
$$H(S_1|S_2) + H(S_2|S_1) < I(W_1, W_2; Y_3|W_0) + I(X_1, X_2; Y_3|W_0, W_1, W_2), \tag{4f}$$
$$H(S_1, S_2) < I(X_1, X_2; Y_3). \tag{4g}$$

We consider all possible error combinations. Assuming that $(i, j, k)$ were sent, (4a) guarantees that the $\Pr(\hat{i} \neq i, \hat{j} = j, \hat{k} = k) < \epsilon$ for any $\epsilon > 0$. (4b) guarantees that $\Pr(\hat{i} = i, \hat{j} \neq j, \hat{k} = k) < \epsilon$, (4c) guarantees that $\Pr(\hat{i} = i, \hat{j} = j, \hat{k} \neq k) < \epsilon$, (4d) guarantees that $\Pr(\hat{i} \neq i, \hat{j} \neq j, \hat{k} = k) < \epsilon$, (4e) guarantees that $\Pr(\hat{i} \neq i, \hat{j} = j, \hat{k} \neq k) < \epsilon$, (4f) guarantees that $\Pr(\hat{i} = i, \hat{j} \neq j, \hat{k} \neq k) < \epsilon$, and (4g) guarantees that $\Pr(\hat{i} \neq i, \hat{j} \neq j, \hat{k} \neq k) < \epsilon$.

The total probability of error can be bounded for large $n$ if (2), (3), and (4a) to (4g) hold. Hence, we have Theorem 1.

We note that in our derivation, we use a correlation structure with a common part for clearer illustration. However, the analysis can be generalized to the case where there is no common part, and hence Theorem 1 is applicable to sources with any arbitrary correlation structure.

# 3 Full Decoding at Destination

Now, we study the strategy when full decoding only occurs at the decoder. First, source coding is performed at each individual source node to remove the correlation among the signals at the nodes. Then we apply channel coding for the MACF to transmit information from the independent sources to the destination.

## 3.1 Source Coding for Correlated Sources

First, we consider a noiseless channel. With node 1 knowing only $s_1$, node 2 knowing only $s_2$, the destination can reconstruct $(s_1, s_2)$ reliably if node 1 encodes $s_1$ with rate $R_1$ and node 2 encodes $s_2$ with rate $R_2$ [1], where

$$R_1 \geq H(S_1|S_2), \tag{5a}$$
$$R_2 \geq H(S_2|S_1), \tag{5b}$$
$$R_1 + R_2 \geq H(S_1, S_2). \tag{5c}$$

## 3.2 Combine with Partial Decode-Forward for MACF

An achievable rate region for the MACFCS can be derived by combining the source coding rate constraints ((5a)-(5c) in Section 3.1) and the channel coding constraints for the MACF ((3a), (3b), (7a)-(7q) in [4]). We call the strategy used in [4] the partial decode-forward strategy. The proof that this rate region is achievable is straightforward.

## 3.3 Combine with Compress-Forward for MACF

In this section, we derive an achievable rate for the MACF using the compress-forward strategy. Combining this with the source coding rate constraints in Section 3.1, we derive another achievable rate region for the MACFCS.

Using the compress-forward strategy, each node transmits independent information as well as a quantized and compressed version of its received signal. Referring to Figure 3, $j$ and $k$ are independent information after source coding. Consider node 1 as an example first. From the received signal $Y_1$, it produces a quantized version $\tilde{Y}_1$. It then compresses $\tilde{Y}_1$ to $U_1$. In the next block, it sends new information $j$ as well as $U_1$. We can view this as node 1 helping node 2 to send a noisy, quantized, and compressed version of node 2's signal, $k$, without needing to fully decode $k$. Node 2 does likewise.

### 3.3.1 An Achievable Rate Region

**Theorem 2** *Let $(\mathcal{S}_1 \times \mathcal{S}_2, p(s_1, s_2), \mathcal{X}_1 \times \mathcal{X}_2, p^*(y_1, y_2, y_3|x_1, x_2), \mathcal{Y}_1 \times \mathcal{Y}_2 \times \mathcal{Y}_3)$ be a discrete memoryless three-node MACFCS. The source symbols $(s_1, s_2)$ can be reliably transmitted*

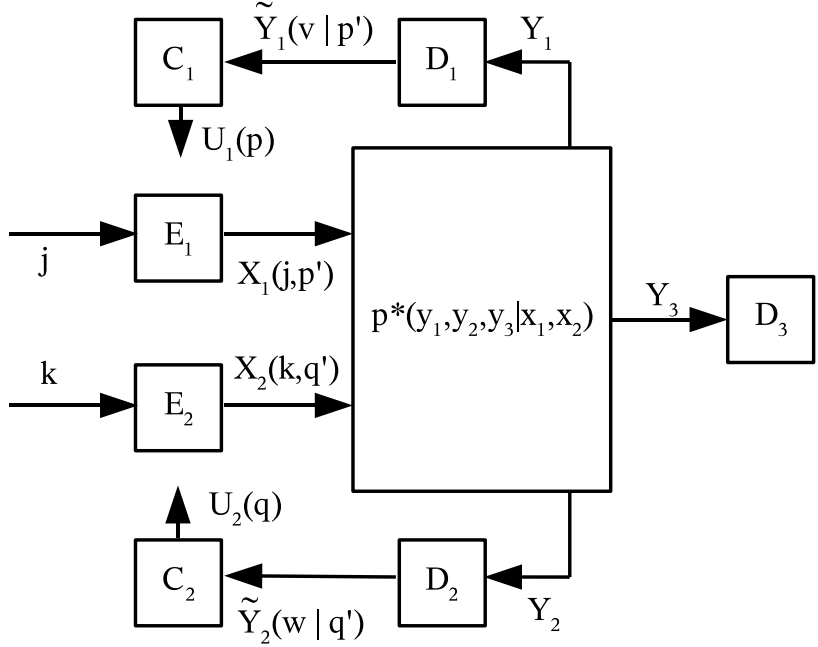

Figure 3: Coding for the multiple access channel with feedback (with independent sources) using the compress-forward strategy.

to the destination if

$$H(S_1|S_2) < I(X_1; \tilde{Y}_2, Y_3|U_1, X_2), \tag{6a}$$
$$H(S_2|S_1) < I(X_2; \tilde{Y}_1, Y_3|U_2, X_1), \tag{6b}$$
$$H(S_1, S_2) < I(X_1, X_2; \tilde{Y}_1, \tilde{Y}_2, Y_3|U_1, U_2), \tag{6c}$$

where the mutual information is taken over all joint probability mass functions $p(u_1)p(x_1|u_1)p(u_2)p(x_2|u_2)p(\tilde{y}_1|y_1, x_1)p(\tilde{y}_2|y_2, x_2)p^*(y_1, y_2, y_3|x_1, x_2)$ such that $\tilde{Y}_1$ and $\tilde{Y}_2$ are independent, subjected to the following constraints

$$I(U_1; Y_3|U_2) > I(\tilde{Y}_1; Y_1|X_1) - I(\tilde{Y}_1; Y_3|\tilde{Y}_2, U_1, U_2), \tag{7a}$$
$$I(U_2; Y_3|U_1) > I(\tilde{Y}_2; Y_2|X_2) - I(\tilde{Y}_2; Y_3|\tilde{Y}_1, U_1, U_2), \tag{7b}$$
$$I(U_1, U_2; Y_3) > I(\tilde{Y}_1; Y_1|X_1) + I(\tilde{Y}_2; Y_2|X_2) - I(\tilde{Y}_1, \tilde{Y}_2; Y_3|U_1, U_2). \tag{7c}$$

Here, $U_1, U_2, \tilde{Y}_1$, and $\tilde{Y}_2$ are auxiliary random variables.

In the next section, we give a brief outline of the proof for Theorem 2.

### 3.3.2 Encoding and Decoding

Figure 3 shows the information data after source coding. Channel encoder 1 receives $j \in \{1, 2, \ldots, 2^{nR_1}\}$ for every $\mathbf{s}_1 = [s_{11}s_{12}\cdots s_{1n}]$. Encoder 2 receives $k \in \{1, 2, \ldots, 2^{nR_2}\}$ for every $\mathbf{s}_2 = [s_{21}s_{22}\cdots s_{2n}]$.

Now, we look at channel coding to ensure that the data bits after source coding can be reliably transmitted to the destination. The codebook generation is as follows.

1. Fix $p(u_1), p(x_1|u_1), p(u_2), p(x_2|u_2), p(\tilde{y}_1|y_1, x_1)$ and $p(\tilde{y}_2|y_2, x_2)$, such that $p(\tilde{y}_1, \tilde{y}_2) = p(\tilde{y}_1)p(\tilde{y}_2)$.

2. Generate $2^{nR'_1}$ i.i.d. sequences $\mathbf{u}_1$ according to $\prod_{i=1}^{n} p(u_{1i})$. Index them $\mathbf{u}_1(p')$, $p' \in \{1, \ldots, 2^{nR'_1}\}$. Generate $2^{nR'_2}$ i.i.d. sequences $\mathbf{u}_2$ according to $\prod_{i=1}^{n} p(u_{2i})$. Index them $\mathbf{u}_2(q')$, $q' \in \{1, \ldots, 2^{nR'_2}\}$.

3. For each $\mathbf{u}_1(p')$, generate $2^{nR_1}$ sequences $\mathbf{x}_1$ according to $\prod_{i=1}^{n} p(x_{1i}|u_{1i}(p'))$. Index them $\mathbf{x}_1(j, p')$, $j \in \{1, \ldots, 2^{nR_1}\}$. For each $\mathbf{u}_2(q')$, generate $2^{nR_2}$ sequences $\mathbf{x}_2$ according to $\prod_{i=1}^{n} p(x_{2i}|u_{2i}(q'))$. Index them $\mathbf{x}_2(k, q')$, $k \in \{1, \ldots, 2^{nR_2}\}$.

4. For each $\mathbf{x}_1(j, p')$, generate $2^{n\tilde{R}_1}$ sequences $\tilde{\mathbf{y}}_1$ according to $\prod_{i=1}^{n} p(\tilde{y}_{1i}|x_{1i}(j, p'))$. Index them $\tilde{\mathbf{y}}_1(v|j, p')$, $v \in \{1, \ldots, 2^{n\tilde{R}_1}\}$. For each $\mathbf{x}_2(k, q')$, generate $2^{n\tilde{R}_2}$ sequences $\tilde{\mathbf{y}}_2$ according to $\prod_{i=1}^{n} p(\tilde{y}_{2i}|x_{2i}(k, q'))$. Index them $\tilde{\mathbf{y}}_2(w|k, q')$, $w \in \{1, \ldots, 2^{n\tilde{R}_2}\}$.

5. Randomly partition the set $\{1, 2, \ldots, 2^{n\tilde{R}_1}\}$ into $2^{nR'_1}$ cells $S_p$, $p \in \{1, \ldots, 2^{nR'_1}\}$; and partition the set $\{1, \ldots, 2^{n\tilde{R}_2}\}$ into $2^{nR'_2}$ cells $S_q$, $q \in \{1, \ldots, 2^{nR'_2}\}$.

The encoding steps are as follows. Basically, node 1 quantizes its received signal from the previous block and compresses it. It sends the compressed information together with its new signal in the new block. Node 2 does likewise.

1. In the beginning of block $t$, remembering its previous transmission in block $t-1$, $\mathbf{x}_1(j^{t-1}, q^{t-2})$; and observing its received signal in block $t-1$, $\mathbf{y}_1(t-1)$, it finds a unique $v^{t-1}$ for which $(\mathbf{x}_1(j^{t-1}, p^{t-2}), \mathbf{y}_1(t-1), \tilde{\mathbf{y}}_1(v^{t-1}|j^{t-1}, p^{t-2})) \in \mathcal{A}_\epsilon$. Using lemma 2.1.3 in [8], node 1 can find such a $v^{t-1}$ with probability tending to 1, with a large enough $n$, if
$$\tilde{R}_1 > I(\tilde{Y}_1; Y_1|X_1). \tag{8}$$
Here, $v^{t-1}$ is the quantized version of $\mathbf{y}_1(t-1)$.

2. Now, node 1 compresses $v^{t-1}$ to $p^{t-1}$. It finds $p^{t-1}$ for which $v^{t-1} \in S_{p^{t-1}}$. It then sends $\mathbf{x}_1(j^t, p^{t-1})$ in block $t$, where $j^t$ is the new message from the source. Here, $p^{t-1}$ is to be decoded and used by the destination to estimate $v^{t-1}$. We see here that node 1 helps node 2 to send a noisy, quantized, and compressed version of node 2's signal to the destination.

3. In block $t$, node 2 quantizes $\mathbf{y}_2(t-1)$ to $w^{t-1}$. It can find a unique $w^{t-1}$ with probability tending to 1 if
$$\tilde{R}_2 > I(\tilde{Y}_2; Y_2|X_2). \tag{9}$$
It compresses $w^{t-1}$ to $q^{t-1}$, where $w^{t-1} \in S_{q^{t-1}}$. It then sends $\mathbf{x}_2(k^t, q^{t-1})$ in block $t$, where $k^t$ is the new information.

The decoding steps are as follows. The destination first decodes the compressed information from nodes 1 and 2. It then estimates the quantized information of the nodes. Using its received signal and the estimated quantized information, it decodes the messages from nodes 1 and 2.

1. At the end of block $t+1$, the destination receives $\mathbf{y}_3(t+1)$. It declares $(\hat{p}^t, \hat{q}^t)$ were sent by nodes 1 and 2 if it can find a unique pair of $(\hat{p}^t, \hat{q}^t)$ for which $(\mathbf{u}_1(\hat{p}^t), \mathbf{u}_2(\hat{q}^t), \mathbf{y}_3(t+1)) \in \mathcal{A}_\epsilon$. This can be done with an arbitrarily small error probability if the following inequalities hold.

$$R'_1 < I(U_1; Y_3|U_2), \tag{10a}$$
$$R'_2 < I(U_2; Y_3|U_1), \tag{10b}$$
$$R'_1 + R'_2 < I(U_1, U_2; Y_3). \tag{10c}$$

2. At the end of block $t$, assume that the destination has correctly decoded $(p^{t-1}, q^{t-1})$ and $(p^t, q^t)$. It find a set $\mathcal{L}(t)$ of $(v^t, w^t)$ such that $\left(\tilde{\mathbf{y}}_1(v^t|j, p^{t-1}), \tilde{\mathbf{y}}_2(w^t|k, q^{t-1}), \mathbf{u}_1(p^{t-1}), \mathbf{u}_2(q^{t-1}), \mathbf{y}_3(t)\right) \in \mathcal{A}_\epsilon$.
It declares that $(\hat{v}^t, \hat{w}^t)$ were sent if it can find a unique $(\hat{v}^t, \hat{w}^t) \in \{(\hat{v}^t, \hat{w}^t) : \hat{v}^t \in S_{p^t} \text{ and } \hat{w}^t \in S_{q^t}\} \cap \mathcal{L}(t)$. This can be done reliably if

$$\tilde{R}_1 < I(\tilde{Y}_1; Y_3|\tilde{Y}_2, U_1, U_2) + R'_1, \tag{11a}$$
$$\tilde{R}_2 < I(\tilde{Y}_2; Y_3|\tilde{Y}_1, U_1, U_2) + R'_2, \tag{11b}$$
$$\tilde{R}_1 + \tilde{R}_2 < I(\tilde{Y}_1, \tilde{Y}_2; Y_3|U_1, U_2) + R'_1 + R'_2. \tag{11c}$$

3. At the end of block $t$, assume that the destination has correctly decoded $(v^t, w^t)$ and $(p^{t-1}, q^{t-1})$. It uses $\tilde{\mathbf{y}}_1(v^t|p^{t-1})$, $\tilde{\mathbf{y}}_2(w^t|q^{t-1})$, and $\mathbf{y}_3(t)$. It declares $(\hat{j}, \hat{k})$ were sent if
$\left(\mathbf{x}_1(\hat{j}^t, p^{t-1}), \mathbf{x}_2(\hat{k}^t, q^{t-1}), \mathbf{u}_1(p^{t-1}), \mathbf{u}_2(q^{t-1}), \tilde{\mathbf{y}}_1(v^t|j, p^{t-1}), \tilde{\mathbf{y}}_2(\hat{w}^t|k, q^{t-1}), \mathbf{y}_3(t)\right) \in \mathcal{A}_\epsilon$.

This can be done with diminishing error probability if

$$R_1 < I(X_1; \tilde{Y}_2, Y_3|U_1, X_2), \tag{12a}$$
$$R_2 < I(X_2; \tilde{Y}_1, Y_3|U_2, X_1), \tag{12b}$$
$$R_1 + R_2 < I(X_1, X_2; \tilde{Y}_1, \tilde{Y}_2, Y_3|U_1, U_2). \tag{12c}$$

Combining these rate constraints for the MACF using the compress-forward strategy and the constraints for the source coding, (5a)-(5c), we get Theorem 2.

## 4  Conclusion

In this paper, we have found a new achievable rate region for the MACF and two new achievable rate regions for the MACFCS. The former is applicable to cooperative wireless communications while the latter is motivated by wireless sensor networks.